\documentclass[letterpaper,aps,prd,preprint,showpacs,nofootinbib,superscriptaddress]{revtex4}
\pdfoutput=1
\usepackage{epsf}
\usepackage{epsfig}
\usepackage{graphics}
\usepackage{graphicx}
\usepackage{amssymb}
\usepackage{amsmath}

\newcommand{\nc}{\newcommand}
\nc{\postscript}[2] 
{\setlength{\epsfxsize}{#2\hsize}\centerline{\epsfbox{#1}}}
\nc{\non}{\nonumber}
\nc{\hc}{\hbox {h.c.}} \nc{\re}{\hbox {Re}} 
\nc{\mev}{\hbox {MeV}} \nc{\gev}{\;\hbox {GeV}} \nc{\tev}{\;\hbox {TeV}}
\def\lsim{\mathrel{\raise.3ex\hbox{$<$\kern-.75em\lower1ex\hbox{$\sim$}}}}
\def\gsim{\mathrel{\raise.3ex\hbox{$>$\kern-.75em\lower1ex\hbox{$\sim$}}}}

\nc{\etal}{{\it et al.}}
\nc{\Lsp}{\;\;\;\;\;\;\;\;\;\;}  \nc{\LLLsp}{\lspace \lspace}
\nc{\lsp}{\;\;\;\;\;\;}
\nc{\spac}{\;\;\;}
\nc{\noi}{\noindent}
\nc{\beq}{\begin{equation}}   \nc{\eeq}{\end{equation}}
\nc{\bea}{\begin{eqnarray}}   \nc{\eea}{\end{eqnarray}}
\nc{\baa}{\begin{array}}      \nc{\eaa}{\end{array}}
\nc{\bit}{\begin{itemize}}    \nc{\eit}{\end{itemize}}
\nc{\ben}{\begin{enumerate}}  \nc{\een}{\end{enumerate}}
\nc{\bce}{\begin{center}}     \nc{\ece}{\end{center}}

\def\sq2{\sqrt{2}}

\def\ph{\varphi}

\def\m4{m^4(\ph)}
\def\mn2{m_n^2}

\def\v5{V^{(5)}}

\begin{document}

\title{\begin{flushright}
       \mbox{\normalsize \rm UMD-PP-08-017}\\
       \mbox{\normalsize \rm NSF-KITP-08-126}
       \end{flushright}
  \vskip 20pt\bf Higgs-Radion Mixing with Enhanced Di-Photon Signal}
\author{Manuel Toharia\\
{\it Department of Physics, University of Maryland\\
College Park, MD 20742, USA}}
\date{\today}

\begin{abstract}
In the context of warped scenarios in which Standard Model (SM)
fields are allowed to propagate in the bulk, we revisit the possible
mixing between the IR localized Higgs field and the Radion graviscalar. 
The phenomenology of the resulting mostly-Higgs field does
not suffer important deviations with respect to the case in which all
the SM is localized in the IR brane (original Higgs-Radion mixing scenario).
On the contrary, the phenomenology of the mostly-Radion field
can present important differences with respect to the original
scenario. At the LHC, the most striking effect is now the possibility
of sizeable Radion decays into photons in a mass range well beyond the
ZZ and WW thresholds, not due to dramatically enhanced couplings to
photons but to suppressed couplings to massive fields.

\end{abstract}

\maketitle

\section{Introduction}
Models of warped extra dimensions have generated a lot of attention as
an interesting and novel framework in which to address simoultaneously
issues such as the hierarchy problem and fermion mass hierarchies. 
In the original setup of Randall and Sundrum \cite{RS1}, the hierarchy problem is dealt with by
localizing all the SM particles in the IR brane. We will refer to this
original proposal as RS1 when comparing it with more recent proposals in
which SM fields are allowed to propagate in the bulk, which will be
loosely referred to as Bulk Fields scenarios.

Precision tests from Electroweak observables put strong bounds on
extensions of the Standard Model (SM), and in the case of the RS1
proposal one should expect higher dimension operators of IR fields to
contribute too importantly to these observables given that the IR
cutoff scale is in the TeV region. By allowing gauge fields (of a
generically extended gauge group) and fermions to propagate in the
bulk one might effectively suppresses the contribution of higher dimension
operators containing the SM fields
\cite{bulkgauge,nobu,RSeff,ADMS,higgsless,holoHiggs,CGHST,GN,GP,HuberShafi}. 
The Higgs field should be close to the IR boundary if the scenario is to address the
hierarchy problem, and now as a benefit of having bulk fermions, one
can use their geographical location to explain the wide differences in
their couplings with the IR localized Higgs.
But precision electroweak tests as well as tight bounds from
flavor physics will now force the new KK modes of the bulk fields to be
heavier than a few TeV
\cite{bulkgauge,nobu,RSeff,ADMS,higgsless,holoHiggs,CGHST,GN,GP,HuberShafi,Burdman,Huber,Agashe,Weiler,Davidson}. 

In all the previous scenarios there is one scalar which, like the Higgs
must be located near the IR brane. It is a scalar mode of the 5D
gravitational fluctuations, parametrizing a vibration mode
of the inter-brane proper distance, the Radion \cite{RS1,GW,Garriga,Rubakov}.
The original setup predicts its mass to be zero due to the fact
that the actual inter-brane distance is not fixed by the spacetime
background setup. Much research has been devoted to study the
stabilization of this setup and therefore the mechanism to generate a
Radion mass \cite{GW,CGRT,TanakaMontes,Pospelov,CGK}. The phenomenological interest in the Radion
lies in the fact that its interactions with SM matter are TeV scale
and therefore could be observable in high energy
collisions \cite{GRW,Kingman,RizzoHewett,Gunion,GTW,Korean}. Moreover its mass remains more or
less a free parameter, and with the assumption that the stabilization
mechanism does not alter importantly the gravitational background, the Radion
is expected to be a much lighter field than the rest of KK excitations \cite{CGK}.

As already mentioned, the Radion is located near the IR Brane and its
interactions are generically proportional to the mass of
the fields it couples to (it couples through the trace of the energy
momentum tensor). These same attributes are shared by the Higgs scalar
and so it becomes important to study carefully the phenomenology
of the full scalar sector to understand how to distinguish between
fields. As it turns out the Higgs and Radion can also mix through a
gravitational kinetic mixing term \cite{GRW} opening the door to even more
interesting consequences \cite{CGK,RizzoHewett,Gunion,GTW}.
Surprisingly most of the research related to the Radion was limited to
the RS1 scenario (except for \cite{Rizzo}), and only recently this research
gap was addressed \cite{CHL}. We plan to extend the study of Radion phenomenology when
Fields are in the Bulk by allowing for the possibility of Higgs-Radion
mixing. In Section II we will review the Radion setup and its
couplings before any Higgs-Radion mixing, which will then be
introduced. In section III we will compare the phenomenology of the
Radion between the RS1 scenario and the Fields in the Bulk models.
The most striking difference will lie in the di-photon channel and we
will concentrate our attention mostly on this channel. Finally in
section IV we will present our conclusions.

\section{Setup}
The spacetime structure consists of one extra dimension with warping such that the
metric takes the usual Randall-Sundrum form \cite{RS1}:
\bea
ds^2=e^{-2\sigma} \eta_{\mu\nu} dx^\mu dx^\nu -dy^2 \label{RS}
\eea 
where $\sigma(y)=ky$, and $k$ is the 5D curvature.
This is a 4D Poincare invariant metric solution in a 5D setup with
bulk cosmological constant and fine-tuned brane tensions at the two
boundaries $y=0$ and $y=y_{{}_{IR}}$ of the extra dimension.

We assume that the origin of Electroweak Symmetry Breaking (EWSB) is
localized in the TeV brane and is well described by a Higgs
doublet. The SM fermions do have profiles along the extra dimension but their
couplings with the Higgs are also localized in the IR brane, and will
depend on the value of their wave-functions at that boundary.

In the gravitational sector, one needs to add full 5D tensor
perturbations $h_{{}_{AB}}(x,y)$ around the metric background
$g_{{}_{AB}}^{RS}$ of Eq.~(\ref{RS}), i.e.
\bea
g_{{}_{AB}}=g_{{}_{AB}}^{RS}+ \hat{\kappa}\ h_{{}_{AB}}
\eea
where $\hat{\kappa}$ is a small parameter.

Thanks to the 5D diffeomorphism invariance we can reduce some of the
linear metric perturbation degrees of freedom to obtain the simpler
perturbative metric \cite{Rubakov,Toharia}:
\bea
\hspace{-1cm} ds^2 &=& \left(e^{-2\sigma} \left[ \eta_{\mu\nu}\ +
  \hat{\kappa} h^{TT}_{\mu\nu}(x,y) \right] - 
\hat{\kappa}\ \eta_{\mu\nu}  { r(x)} \right)  dx^\mu dx^\nu -  \left(1 + \hat{\kappa}\  2 e^{2\sigma}{ r(x)} \right)  dy^2 \label{metricpert}
\eea
where $h^{TT}_{\mu\nu}(x,y)$ is transverse and traceless and $r(x)$ is
the Radion graviscalar which cannot be gauged away due to the presence
of the two brane boundaries. In the absence of a stabilization
mechanism the Radion is massless (and therefore a
problematic long-range force mediator), but it was quickly
realized that a very simple fix to this was to add an extra bulk
scalar field to the setup, such that it acquires a nontrivial
background vev along the extra dimension \cite{GW}. This space-time background
solution will fix the inter-brane distance and generically give a
positive mass squared to the Radion. The stabilized background metric 
solution can be in some limit very close to the Randall-Sundrum solution, so that we
can still use Eq.~(\ref{RS}) as the background metric. We will refer to this limit as the ``small
back-reaction limit'' and will assume it for the remaining of the
paper. We will consider the radion mass as a free parameter although
we should expect it to be relatively light in the small back-reaction limit, at
least when invoking a Goldberger-Wise type stabilization mechanism \cite{CGK}.

\subsection{Interactions}

From the perturbative ansatz of Eq.~(\ref{metricpert}), it is simple to extract
the tree-level interactions between the Radion and the matter fields since these
are just gravitational interactions. One obtains the linear Radion-matter interactions as \cite{Rizzo,CHL}
 \bea
\hspace{-2cm}S_1^{int}\!\! &=&\!\! - {\hat{\kappa} \over 2}\   \int d^5x \ e^{-2\sigma}
\left(-T^\mu_{\ \mu}+2 T_{5 5}  \right) \ \ r(x)  \label{radtraceT}
\eea
where the graviscalar field $r(x)$ is not a 4D canonically normalized
scalar field. The physical Radion field $\phi_0(x)$ is obtained with the redefinition 
\beq
{\hat{\kappa}\over2} r(x)=-{1\over \Lambda_\phi} \phi_0(x).
\eeq 
where $ \Lambda_\phi = \sqrt{6}\ M_{Pl}\ e^{-k y_{{}_{IR}}}$ and the minus sign restores the 
convention of \cite{Gunion}. The 5D matter field information is
included in $T_{AB}$, the energy momentum tensor.
From the previous tree-level interactions one needs to extract the interactions
between the Radion and the lightest modes of the 5D bulk matter, 
i.e the SM massive gauge bosons and fermions.
Assuming that the 5D electroweak group is simply $SU(2)\times U(1)$,
and that the fermion structure is a simple 5D extension of the
Standard Model with a Higgs mechanism localized on the IR brane, the
couplings to the Radion are \cite{CHL}
\bea
&&{M^2_V} \left(1-6\ k y_{{}_{IR}}{M^2_V\over\Lambda_\phi^2} \right)\   {\phi_0\over \Lambda_\phi} V^\alpha V_\alpha, \label{rVV}  \\
&&{ m_f}(c_L-c_R)\ {\phi_0\over \Lambda_\phi}\bar{f}_{UV}f_{UV},\label{rff1}\\
&&{ m_f}\ {\phi_0\over \Lambda_\phi}\bar{f}_{IR}f_{IR}\label{rff2}
\eea
where $f_{UV}$ and $f_{IR}$ represent fermions localized near the UV
and IR brane respectively, with $c_L$ and $c_R$ being the mass
parameters associated to the left and right handed bulk fermions.
The term proportional to $ky_{{}_{IR}}$ in the massive boson couplings and the term proportional to the fermion
mass parameters $c_i$ appearing in the fermion couplings are new
contributions due to the 5D nature of matter fields, and were not
present in the RS1 scenario.

In the case of massless gauge bosons, i.e. gluons and photons, the
interactions with the Radion appear with same strength from various
sources \cite{CHL}. First, there is a one-loop contribution identical to the Higgs radiative
couplings; also, because the gauge interactions are 5-dimensional there is a
tree-level interaction with photons which can be directly extracted
from Eq.~(\ref{radtraceT}). Brane kinetic terms associated with the gauge fields will
also contribute, if present, and finally there is a term proportional
to the total gauge group beta function coefficient, coming from the trace
anomaly for IR light fields and from loop corrections to the gauge
coupling due to UV and bulk fields. We write 
\bea
\left[{1-4 \pi  \alpha (\tau_{UV}^0+\tau_{IR}^0)\over 4\ k y_{{}_{IR}}}
+ {\alpha\over 8 \pi}\left(b-\sum_i \kappa_i F_i(\tau_i)\right)
\right]\ {{ \phi_0}\over \Lambda_\phi} F_{\mu\nu}F^{\mu\nu}.\label{rgg}
\eea
where $\tau_{UV}^0$ and $\tau_{IR}^0$ are the brane kinetic terms, $\sum_i \kappa_iF_i$
are the contributions from the one-loop diagrams and $b$ is the total
beta function coefficient associated with the corresponding gauge
field. The first term in this formula is a consequence of the
5-dimensional nature of the gauge fields and again was not present in
the RS1 scenario.
\begin{figure}[t]
\center
\includegraphics[width=10cm]{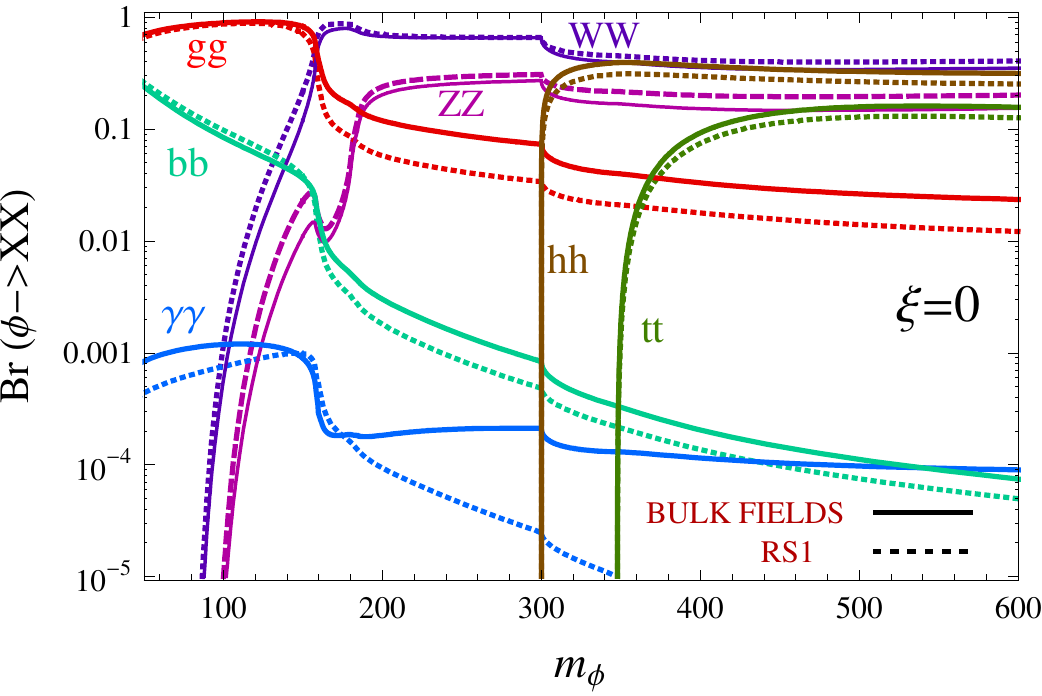}
\caption{Branching fractions for the Radion as a function of its mass $m_\phi$
  in the RS1 scenario (dotted curves) and the Fields in the Bulk scenario
  (thick curves). The individual curves are very similar for both
  scenarios except for the $\gamma\gamma$ channel (blue curve) which
  drops quickly with the Radion mass in the RS1 case but becomes flat
  in the Fields in the Bulk case 
}
\hspace{.1cm}
\label{fig1}
\end{figure}

In Fig.~\ref{fig1} we show the branchings of the Radion in the
absence of Higgs-Radion mixing for the two scenarios we wish to compare,
RS1 and the Fields in the Bulk. In
this last case we have not included any brane kinetic term associated
with the bulk (gauge) fields and we will not include them either in
the rest of this study\footnote{Brane kinetic terms for photons and/or
  gluons will actually have interesting effects in both suppressing or
  enhancing the $gluon$-$gluon$ and the $\gamma\gamma$ decays as shown in
  \cite{CHL}.}. The figure agrees
reasonably well with \cite{CHL}, the main difference being
that we have included here decays into one intermediate off-shell $Z$
or $W$ boson, a process which can still be quite significant below the
$ZZ$ and $WW$ physical thresholds. One observes that the branchings do
not vary much between the two scenarios and we mainly point out for
the case of the Fields in the Bulk, a slight increase in $gluon$-$gluon$
branchings and an interesting plateau for the $\gamma\gamma$ branchings
in the high Radion mass range.

\subsection{Higgs-Radion mixing}

One can still modify this setup since it is always possible to write down localized gravity kinetic
terms in the boundaries,
\bea
 S^{BKT}_i=\pm M_i^2 \int d^4x\ \sqrt{-g_i}\ R_i 
\eea
where $g_i$ is the determinant of the induced metric on the $i$-th
brane, and $M_i$ is some dimensionful parameter that should be
naturally related to the fundamental scales $M_5$ and $k$.
These terms will contribute to the kinetic term of
the Radion and their presence will therefore require a new canonical normalization. 
In the same lines, in the IR brane we have
at our disposal the Higgs bilinear $H^+H$ which can be coupled to
the induced Ricci scalar $R$ to obtain an effective dimension-4
operator \cite{GRW},
\bea
 S_\xi=\xi \int d^4x\ \sqrt{-g_{IR}}\ R H^+H 
\eea
where $\xi$ is a dimensionless parameter.

In the small back-reaction limit, and after EWSB, the two lightest (mixed) states of
the scalar sector will be the Radion graviscalar and the Higgs; the
effective 4d action for these fields up to quadratic order will be:
\bea
{\cal L}=-{1\over2}\left\{1\pm 6 {M_{IR}^2\over\Lambda^2_\phi}+6\gamma^2 \xi \right\}\phi_0\Box\phi_0
-{1\over2}\phi_0 m_{\phi_o}^2\phi_0-{1\over2} h_0 (\Box+m_h^2)h_0-6\gamma \xi \phi_0\Box h_0\,,
\label{keform}
\eea
where $m_{ho}^2$ and $m_{\phi_o}^2$ are the Higgs and Radion masses
before mixing and $\gamma= v/\Lambda_{\phi}$ is a dimensionless
parameter reflecting the suppression of the Radion-to-matter couplings
with respect to the Higgs-to-matter couplings.
For simplicity and to make close contact with previous work, we will set
$M_{IR}=0$\footnote{Non-zero $M_{IR}$ will only affect the Radion kinetic term and can be
recast as an overall modification of the ``bare'' Radion couplings to
matter (i.e. a redefinition of $\Lambda_\phi$). The effects of such terms on the KK graviton
spectrum and couplings have been studied in \cite{Rizzobrane}}.

The states $h(x)$ and $\phi(x)$ which diagonalize (\ref{keform}) can be introduced as
\bea
\left(\begin{array}{c}
h_0 \\ \phi_0
\end{array}\right) = 
\left(\begin{array}{cc}
d&c \\ b&a
\end{array}\right)
\left(\begin{array}{c}
h \\ \phi
\end{array}\right)\label{abcd}
\eea
It is interesting to remark that this transformation is not
orthogonal due to the nature of the mixing (kinetic mixing and
extra contribution to the Radion kinetic term).
We can decompose the previous transformation in terms of two
operations. The first redefinition diagonalizes and normalizes the kinetic terms,
and the second one, an orthogonal transformation this time, diagonalizes the
mass matrix. 
To maintain positive definite kinetic energy terms for the Radion $\phi$,
we must have $Z^2>0$, where
\bea
Z^2&\equiv& 1+6\xi\gamma^2(1-6\xi)\,.
\label{z2}
\eea
This allows us to obtain theoretical limits on
the $\xi$ parameter in terms of the scale $\Lambda_\phi$ (i.e. the
parameter $\gamma =v_0/\Lambda_\phi$):
\bea
{1\over12}\left(1-\sqrt{1+ {4\over \gamma^2}}\right) \leq \xi \leq{1\over12}\left(1+\sqrt{1+ {4\over \gamma^2}}\right).
\eea 
The corresponding mass-squared eigenvalues of the new physical states
are~\footnote{Note that the quantity inside the square root is positive definite so long
as $m_{h_o}^2m_{\phi_o}^2>0$.}
\beq
m_\pm^2={1\over 2 Z^2}\left(m_{\phi_o}^2+\beta m_{h_o}^2\pm\sqrt{
[m_{\phi_o}^2+\beta m_{h_o}^2]^2-4Z^2m_{\phi_o}^2m_{h_o}^2}\ \right)\,,
\label{emasses}
\eeq
where we have defined $\beta=1+6\xi \gamma^2$. These masses must
satisfy the inequality
\beq
{m_+^2\over m_-^2}>1+{2\beta\over Z^2}\left(1-{Z^2\over\beta}\right)+{2\beta\over Z^2}\left[1-{Z^2\over \beta}\right]^{1/2}\,,
\label{rootconstraint}
\eeq
in order for the `bare' masses $m_{\phi_o}^2$ and $m_{h_o}^2$ to be
real \cite{Gunion}. This constraint on the physical masses is quite interesting
since the larger the value of $|\xi|$, the larger the mass-squared difference
of the two scalar fields must be.

The parameters needed to fix the scalar sector are the Higgs and Radion
`bare' masses $m_{\phi_o}^2$ and $m_{h_o}^2$ before mixing, the Radion interaction scale $\Lambda_\phi$
(related to the KK masses and the solution of the hierarchy problem) and the mixing
parameter $\xi$. We will trade the `bare' masses of the original fields
for the phenomenologically more interesting physical masses of
the two mixed scalar states. We will refer to these as the Higgs
and the Radion, even though they are an admixture of both, and will
fix the convention by defining the Higgs scalar as the field which
becomes the SM Higgs in the limit of $\xi\to 0$ and similarly for the
Radion. The important parameters are thus
$$m_h,\ \ m_\phi,\ \ \Lambda_\phi\ \  {\rm and}\ \ \xi,$$
with the understanding that they are not completely independent since
the masses are bound by the non-degeneracy constraint from
Eq.~(\ref{rootconstraint}), which defines a theoretically forbidden
region for these parameters.

\section{Phenomenology}

The couplings of the physical scalar fields with the SM matter fields
will be obtained using the redefinitions of Eq.~(\ref{abcd}).
Calling $g_{hii}^0$ and $g_{\phi ii}^0$ the coefficients of the 'bare'
Higgs and Radion couplings to the fields ``$i$'', one obtains 
\bea
g_{hii}&=&d\ g_{hii}^0 +b\ g_{\phi ii}^0\label{hcoupling}\\
g_{\phi ii}&=&c\ g_{h ii}^0 +a\ g_{\phi ii}^0\label{phicoupling}
\eea 
for couplings of the physical fields.

Now, the couplings $g_{\phi ii}^0$ of the 'bare' Radion are basically
the same as the couplings of the `bare' Higgs $g_{h ii}^0$, but
suppressed by a factor of $\ \sim v/\Lambda_\phi$, where $v$ is the Higgs vev.

When the mixing parameter $\xi$ is small, the redefinition coefficients $b$ and
$c$ must be small too (basically they play the role
of $\sin{\theta}$ in the case of a typical orthogonal mass
mixing), whereas $d$ and $a$ must lie close to 1 (like $\cos{\theta}$
for small $\theta$). A quick glance at the previous couplings shows that
for small mixing, the couplings will look like
\bea
g_{hii}&\sim& g_{hii}^0(1 +b {v\over \Lambda_\phi})\label{hcouplingapp}\\
g_{\phi ii}&\sim& g_{h ii}^0(c +{v\over\Lambda_\phi})\label{phicouplingapp}
\eea 
where both $b$ and $c$ are small numbers. One sees that the Higgs couplings
do not receive much corrections, since $b$ is small and is multiplied by
the also small $v/\Lambda_\phi$. Of course for larger values
of the mixing parameter $\xi$, the couplings of the mostly-Higgs
scalar will start deviating significantly from the SM Higgs
values. This effect was extensively studied in \cite{Gunion} in the context of RS1 and we will not pursue 
it further here, but instead concentrate on the mostly-Radion sector were
the couplings are more sensitive in the small mixing
region. Moreover, when the mixing is large one should also consider
carefully the bounds coming from precision electroweak constraints,
as now both the Radion and the Higgs are expected to contribute
importantly to electroweak observables such as S and T \cite{CGK,GTW}.
  
On the other hand, the Radion couplings can quickly
change for small mixing since now, even if $c$ is small, it is to be compared with
$v/\Lambda_\phi$, which is small too (see Eq.~(\ref{phicouplingapp})). In particular this means that
for a small mixing in the appropriate direction, one can actually
suppress completely the physical coupling of the Radion to the fields
``$i$''. Of course, in principle, the point where the
Radion couplings to the ``$i$'' particles vanish does not
mean that the couplings to some different ``$j$'' particles also vanish.

\begin{figure}[t]
\center
\includegraphics[width=8.1cm,height=6cm]{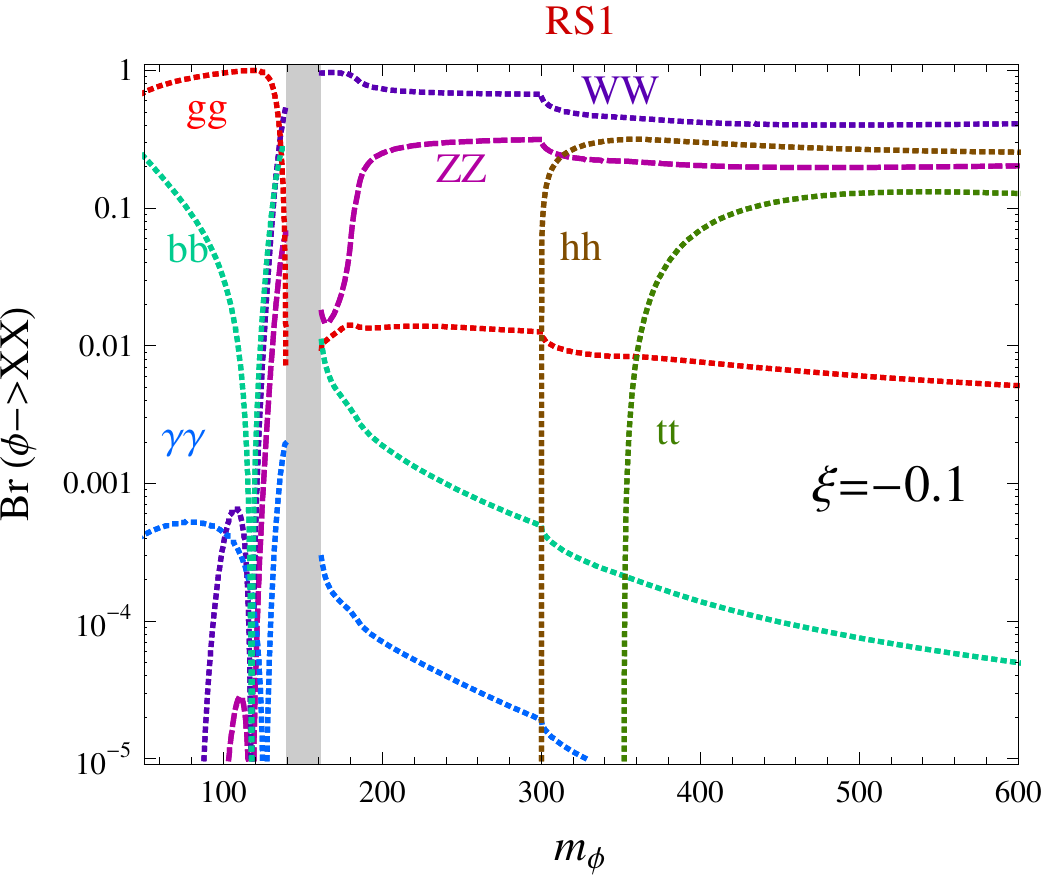}
\includegraphics[width=8.1cm,height=6cm]{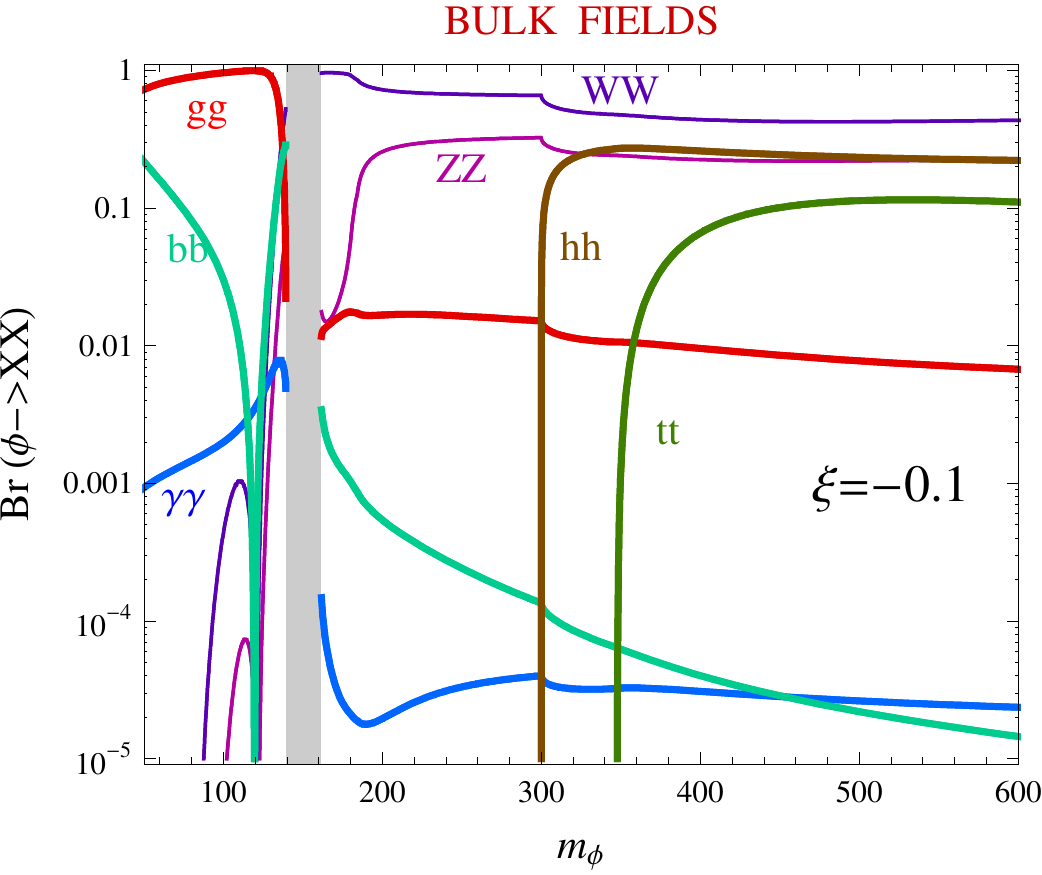}
\caption{Branching fractions for the Radion as a function of its mass $m_\phi$
  in the RS1 scenario (left panel) and the Fields in the Bulk scenario
  (right panel). We have taken a small and negative mixing parameter
  value, $\xi=-0.1$, and a physical Higgs mass of $m_h=150$ GeV.
  In both panels, the vertical gray band centered at the physical Higgs mass 
  represents the theoretically excluded region in which the values of
  the `bare' scalar masses are complex. The individual curves are
  very similar in both panels except for the $\gamma\gamma$ channel
  (blue curve) which shows different features in each scenario
  reaching higher maximum values in the Fields in the Bulk scenario. 
}
\label{fig2}
\end{figure}

\begin{figure}[t]
\center
\includegraphics[width=8.1cm,height=6cm]{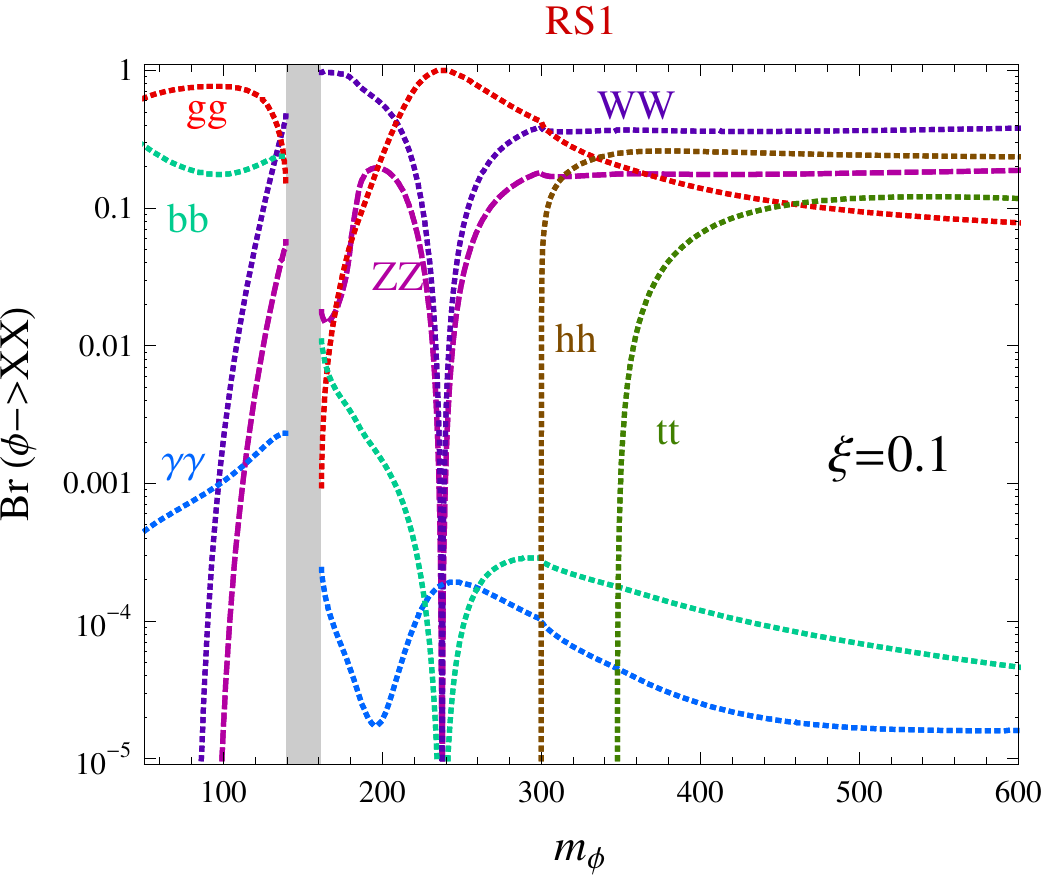}
\includegraphics[width=8.1cm,height=6cm]{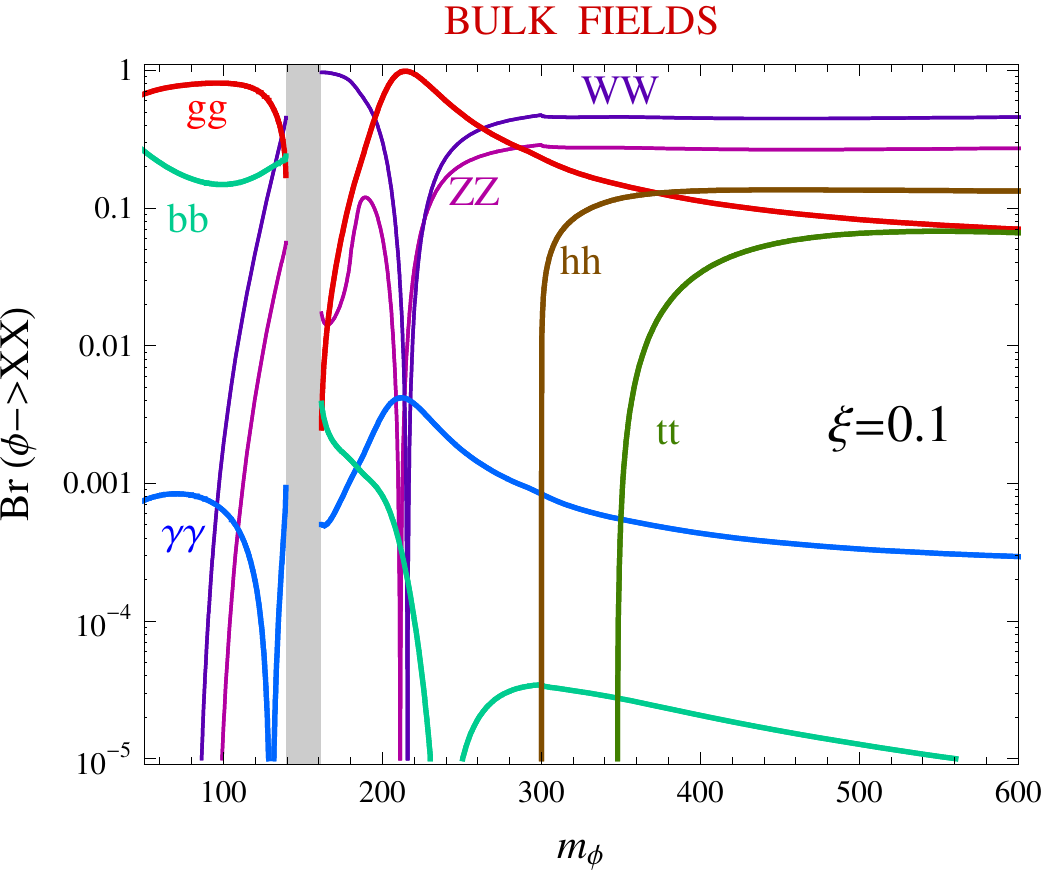}
\caption{Same as Fig.~\ref{fig2} but with $\xi=+0.1$. Again the $\gamma\gamma$ channel
  (blue curve) shows very different features in each scenario, and in
  particular, in the right panel, one observes strikingly high values
  in this channel for a range of Radion masses beyond the ZZ and WW
  threshold. This effect is mainly due to a suppression in the
  couplings of the Radion to massive gauge bosons.}
\label{fig3}
\end{figure}

Nevertheless, in the RS1 scenario in which all fields lie on the IR brane, it turns
out that the point where the physical Radion is phobic to $W$ bosons,
is the same where it is phobic to $Z$ bosons and to all massive fermions. Moreover, the suppressed
couplings to $\gamma\gamma$ happen in a nearby point. These features are a consequence of the exact Higgs-like
structure of the Radion couplings to massive particles in RS1 (vector bosons and
fermions) where by simply replacing the parameter $\Lambda_\phi$ by the Higgs vev $v$, one obtains
the Higgs couplings. Moreover the trace anomaly contribution to the
Radion coupling to $\gamma\gamma$ is numerically small, and therefore
this coupling still resembles the Higgs one. 
In the case of the Radion coupling to gluons the contribution from the
trace anomaly is quite important and therefore the zero for that
coupling is very separated in parameter space from the points where
the other couplings vanish. 

In the left panels of both Fig.~\ref{fig2} and Fig.~\ref{fig3},
we plot the Radion decay branching fractions with respect to its
mass for the RS1 scenario, for two different small mixing parameters,
$\xi=-0.1$ and $\xi=0.1$. We have taken $\Lambda_\phi=2000$ GeV and the mass of the mostly-Higgs
scalar is $m_h=150$ GeV. In both plots one observes that there is a `universal'
point where all the couplings to massive
particles vanish. The $\gamma\gamma$ signal is suppressed in a nearby
region, so one cannot take advantage of the suppression of other
couplings, and the gluon branchings are suppressed only in the
region close to the theoretically forbidden boundary, represented in
the plots as a vertical gray band, centered at the physical Higgs
mass, here $m_h=150$ GeV (The Radion and the Higgs masses cannot lie 
in the same region, as explained below Eq.~(\ref{rootconstraint})).

When the matter fields are placed in the 5D bulk, the couplings of the Radion
receive some corrections relative to the RS1 case, as explained below
Eqs.~(\ref{rVV}-\ref{rff2}). The consequence of this is that now the
zeroes of the different Radion couplings happen in separated points of
parameter space. In the case of the $WW$ coupling and the $ZZ$
coupling this separation is not very substantial, although it does
happen. Couplings to different families and types of fermions will have zeros
in different regions of parameter space, depending on the
values of the mass parameters $c^i_{L,R}$ for each fermion.
But the most important change comes from the new
contribution to the Radion coupling to photons (see Eq.~(\ref{rgg})),
where the sign of the overall coupling actually flips, even if retaining a similar
absolute value. Therefore, when we turn on the $\xi$-mixing, the
coupling of the Radion to photons will vanish in a very different
region where it used to in the RS1 scenario.
This feature is apparent in the right panels of both Fig.~\ref{fig2} and Fig.~\ref{fig3},
where one observes clearly how the suppression to $\gamma\gamma$ happens far
away from the other zeroes (namely on the opposite side of the forbidden
vertical band). Then, the branching fraction to photons
increases substantially in the region where the Radion is phobic to 
massive particles. Specially in the right panel of Fig.~\ref{fig3} we also observe how the zero to
$b\bar{b}$ is moved away from the $WW$ and $ZZ$ zeroes. For this, we
used $\ c_L-c_R=1.1\ $ for the bottom quark 5D mass parameters. One
could study in detail the variations of the couplings to other
fermions such as tau's and charm quarks, since the branchings could
change importantly if one happens to live near a zero coupling for
one of these heavier fermions. At the LHC, generically the Radion is mainly
produced in gluon fusion, with other production mechanisms extremely
suppressed (unlike the SM Higgs case). In that case most fermionic
decays will be hard to study due to the enormous QCD background. When
the Higgs-Radion mixing is large, production via vector boson fusion
for example can be enhanced importantly \cite{Gunion}, therefore opening the door to
study decays into tau's with associated forward and backward jets,
just like in the SM Higgs case. We will not pursue further this line of investigation
although it might be an interesting one for the future.
\begin{figure}[t]
\center
\includegraphics[width=7.cm,height=7cm]{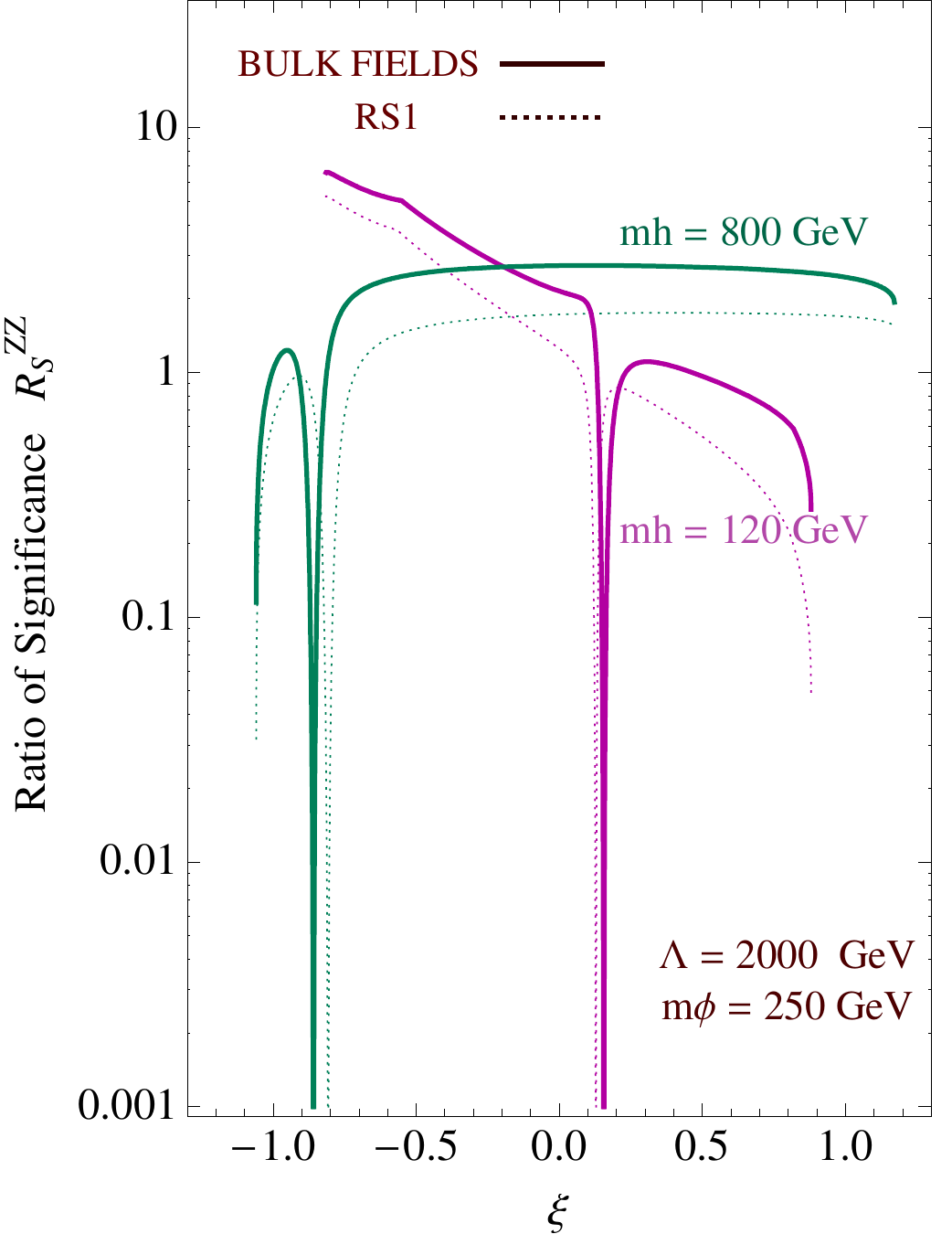}\hspace{1cm}
\includegraphics[width=6.7cm,height=7cm]{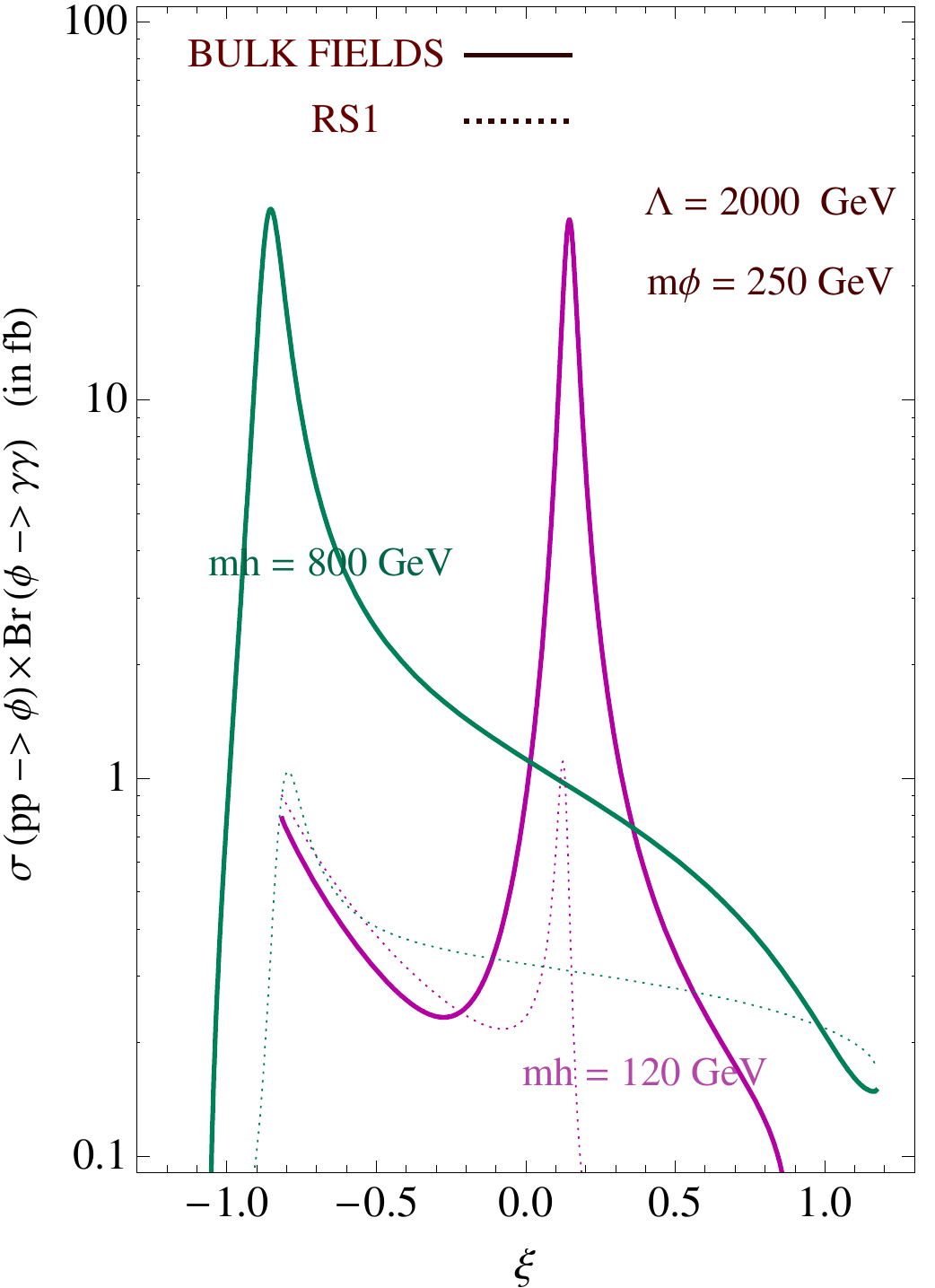}
\caption{{\it Left Panel:} Ratio of discovery significances 
$R_S^{ZZ}={S(gg\to\phi\to ZZ)/S(gg\to h_{SM}\to ZZ)}$ in the $ZZ$ channel between the Radion
and a SM Higgs of same mass as a function of $\xi$.\ \  {\it Right
Panel:} Cross section (in fb) for the process
$pp\to\phi\to\gamma\gamma$ as a function of $\xi$. In both panels we
take $\Lambda_\phi=2$ TeV and the radion mass $m_\phi=250$ GeV, and
the dotted (solid) curves are for the RS1 (Bulk Fields) scenario. We
take two limits for the mass of the mostly-Higgs scalar, $m_h=120$ GeV (Purple) and
$m_h=800$ GeV (Green). The dependance on $\xi$ in the $ZZ$ channel is
similar for both RS1 and Bulk Fields scenarios, whereas the
$\gamma\gamma$ signal fluctuates quite importantly with $\xi$ in the Bulk
Fields scenario.}
\label{fig4}
\end{figure}

Instead we plan to focus on the Radion decays into photons since the
branchings can be substantially enhanced for both light masses,
$110$\ GeV$<m_\phi<150$ GeV, but also more massive ones, $m_\phi>150$
GeV, as shown in the right panel of Fig.~\ref{fig3}. But before
concentrating on the $\gamma\gamma$ channel we want to also take a
look at the Radion decays into $ZZ$, a very important channel for the
large mass region. In \cite{CHL} it was observed that, in the
absence of brane gauge kinetic terms, the $ZZ$ signal is relatively
enhanced with respect to the RS1 scenario. In the presence of
Higgs-Radion mixing one would expect this enhancement to remain
similar. 
This is confirmed in the left panel of Figure \ref{fig4}.
The ratio of discovery significances 
$R_S^{ZZ}={S(gg\to\phi\to ZZ)/S(gg\to h_{SM}\to ZZ)}$, as defined in \cite{GRW,CHL}, in the $ZZ$ channel between the Radion
and a SM Higgs of same mass is plotted as a function of $\xi$. The dotted curves
are for the RS1 scenario, while the solid ones are for the Fields in the Bulk scenario.
We have taken $\Lambda_\phi=2000$ GeV and a Radion mass of $m_\phi=250$ GeV, and chose a
light Higgs scenario ($m_h=120$ GeV) and a heavy Higgs one ($m_h=800$
GeV). In the two cases the RS1 curves and the Bulk Field curves follow
each other closely, as expected.

In the right panel we plot the cross section (in fb) for the process
$pp\to\phi\to\gamma\gamma$ as a function of $\xi$. We have computed it
for $\sqrt{s}=14$ TeV without QCD corrections and used CTEQ5L pdf's.
The difference between the RS1 scenario and the Bulk Fields is quite
striking, and the cross sections can reach up to $20-30$ fb for the
parameters chosen. This is roughly the level of cross sections that
one expects for a SM Higgs with a mass of 120-130 GeV, even though the
Radion mass taken here is quite large, $m_\phi=250$ GeV.

\begin{figure}[t]
\center
\includegraphics[width=8.1cm,height=6cm]{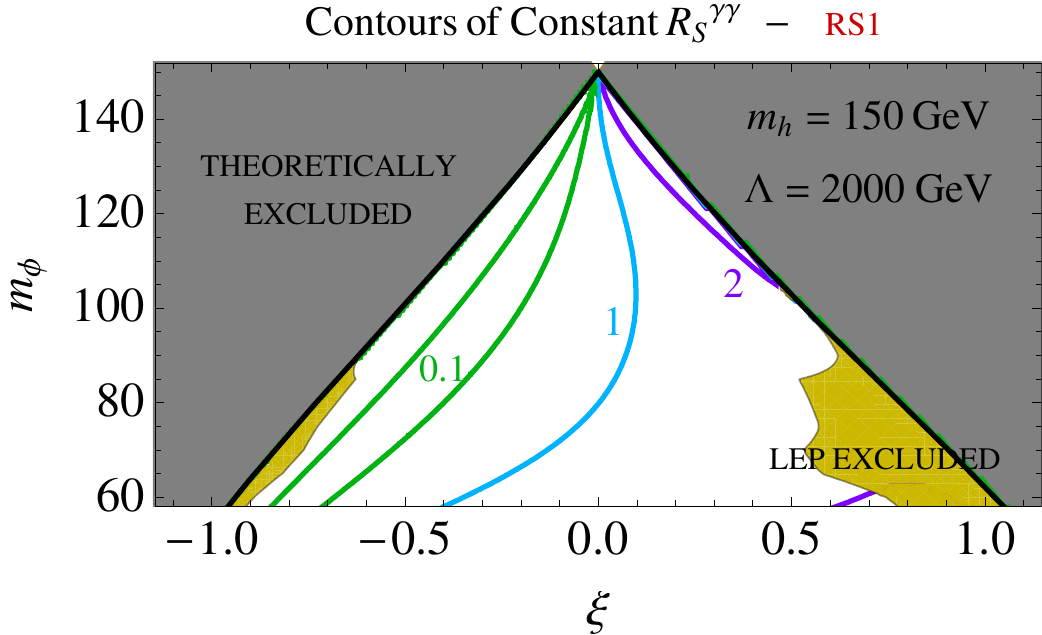}\ \ \includegraphics[width=8.1cm,height=6cm]{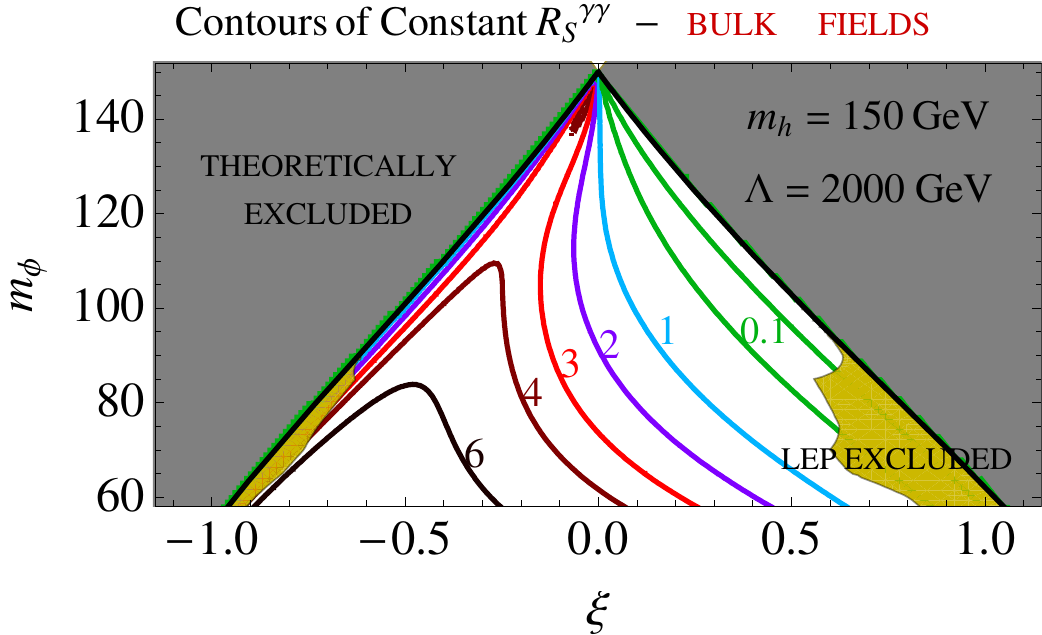}
\caption{Contours in the $(\xi,m_\phi)$ plane of the ratio of discovery significances
 $R_S^{\gamma\gamma}={S(gg\to\phi\to\gamma\gamma)/S(gg\to h_{SM}\to\gamma\gamma)}$, 
with the SM Higgs mass equal to that of the Radion. 
The mass of the $mostly-$Higgs scalar of the setup is held
at $m_h=150$ GeV and the Radion coupling scale at $\Lambda_\phi=2$
TeV. The left panel shows results for the model in which all SM fields
are confined in the TeV Brane (here referred to as RS1) and the right
panel shows results for the case in which all fields propagate in the
bulk except the Higgs, which remains confined in the TeV brane.
}
\label{fig5}
\end{figure}
This prompts us to do a parameter space scan to see how large is the
region where the di-photon signal is important (and enhanced with respect to the RS1
scenario). In Figure \ref{fig5} we choose the mass of the mostly-Higgs
field as $m_h=150$ GeV and the Radion coupling scale at
$\Lambda_\phi=2$ TeV. We then plot, in the $(\xi,m_\phi)$ plane, contours of the ratio of discovery significances
 $R_S^{\gamma\gamma}={S(gg\to\phi\to\gamma\gamma)/S(gg\to
  h_{SM}\to\gamma\gamma)}$, as defined in \cite{GRW,CHL}, between the Radion and a SM Higgs with equal
mass. In RS1 (left panel), for positive values of $\xi$ one can reach roughly up to
2 times better than a SM Higgs. For the Bulk Fields scenario (right panel), the
enhancement happens for negative values of $\xi$, and reaches quite larger
values. This is because the zeroes of the couplings to $W$'s
and $Z$'s happen to be in the negative $\xi$ region while the
zeroes of the $\gamma\gamma$ coupling are now in positive $\xi$
region. In the mass region considered for this figure (below $WW$ and $ZZ$ thresholds),
and at the points of larger di-photon signal, the collider phenomenology would be quite similar to exotic
Higgs scenarios with enhanced di-photon branchings (see for e.g. \cite{Mrenna}).

The gray region is the theoretically excluded region while the
yellow regions are parameter points excluded by LEP data \cite{LEP}, given that
the Radion couplings to $ZZ$ are quite enhanced in those corners
thereby reaching the bounds set by LEP, as previously remarked in \cite{Gunion}.

\begin{figure}[t]
\center
\includegraphics[width=8.1cm,height=6cm]{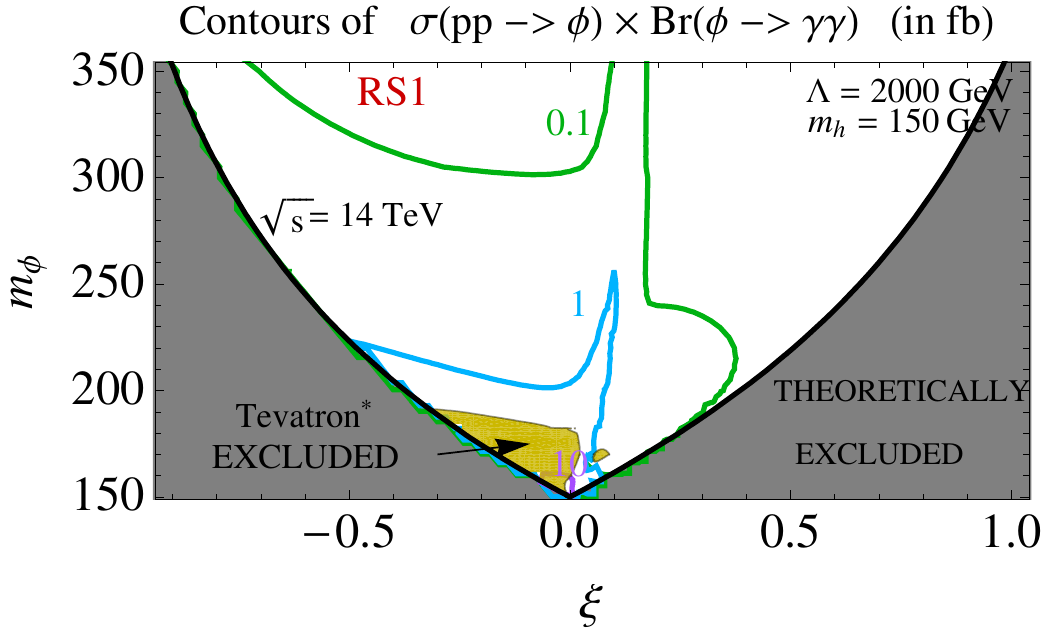}\ \ 
\includegraphics[width=8.1cm,height=6cm]{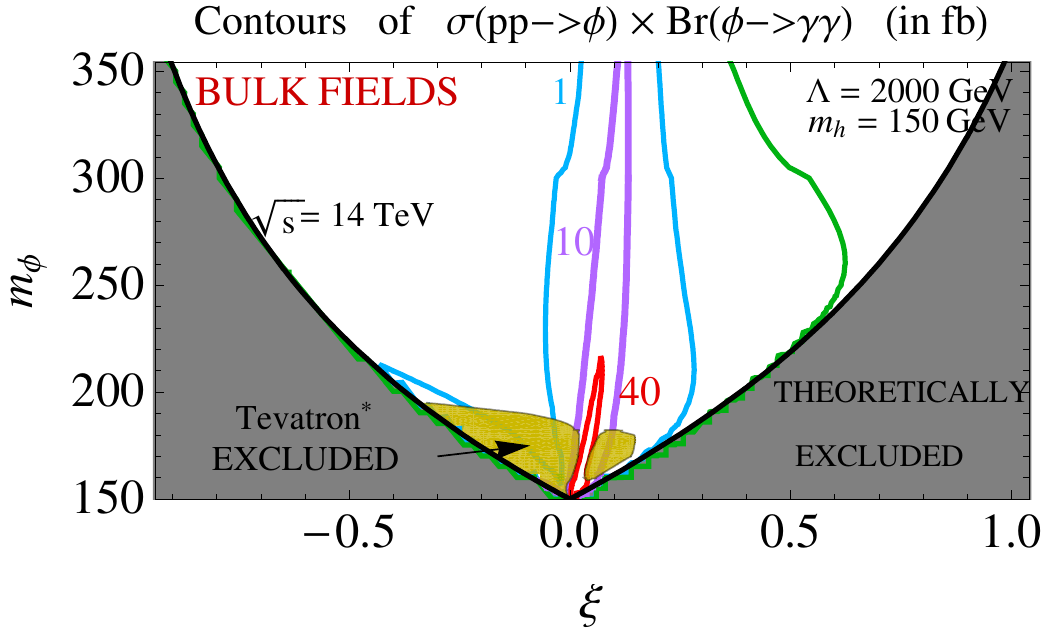}
\caption{Contours in the $(\xi,m_\phi)$ plane of the di-photon cross
  section $\sigma(pp\to\phi)Br(\phi\to\gamma\gamma)$ (in fb) 
at the LHC with $\sqrt{s}=14$ TeV (using CTEQ5L gluon PDF).  
The mass of the $mostly-$Higgs scalar of the setup is held
at $m_h=150$ GeV and the Radion coupling scale at $\Lambda_\phi=2$
TeV. The left panel shows results for the model in which all SM fields
are confined in the TeV Brane (here referred to as RS1) and the right
panel shows results for the case in which all fields propagate in the
bulk except the Higgs, which remains confined in the TeV brane.
}
\label{fig6}
\end{figure}

Surprisingly one can still obtain interesting rates in the di-photon
channel for a heavier Radion. Of course a comparison with a SM Higgs
of same mass becomes useless since this channel is extremely
suppressed above $\sim 160$ GeV. Therefore for the large Radion mass
range, we decide to plot in both panels of Figure \ref{fig6} contours in the
$(\xi,m_\phi)$ plane of the di-photon cross section
$\sigma(pp\to\phi)Br(\phi\to\gamma\gamma)$ (in fb) at the LHC with
$\sqrt{s}=14$ TeV (using CTEQ5L gluon PDF). We have not included QCD
corrections, which tend to enhance the production cross section, but
these are the same for the SM Higgs process and one can easily
estimate their effect. In any case our main interest lies in comparing
different scenarios so we can confidently use the leading order
results.
As in Figure \ref{fig5} we take $m_h=150$ GeV and
$\Lambda_\phi=2$ TeV, and again we observe a striking difference between
the RS1 scenario (left panel) and the Bulk Fields scenario (right
panel). In the RS1 case the cross section above $200$ GeV is at most 1
fb making this signal quite difficult at least in the first years of
running. On the other hand, when fields are in the Bulk, one sees that
there is a thin tower region in the allowed parameter space where
large cross sections are possible. When we calculate the cross section 
$\sigma(pp\to h_{SM})Br(h_{sM}\to\gamma\gamma)$ for a SM Higgs with
mass $m_{h_{SM}}=130$ GeV we find it to lie at around $40$ fb which
is therefore comparable to the red contour (the shortest vertical
band) which reaches Radion masses of about $220$ GeV. 
A cross section of $10$ fb is possible in a much larger region and can
actually be reached for Radion masses
well above $350$ GeV. We should note here that the dependence of the
cross sections with the scale $\Lambda_\phi$ goes as
$\sim 1/\Lambda_\phi^2$, and so for example the numbers in the
contours should be roughly divided by 4 if we were to take
$\Lambda_\phi=4$ TeV and scan for the new allowed regions in the
$(m_\phi-\xi)$ plane. In any case the larger the mass, the
better the backgrounds are, so one should definitely consider this
surprising channel as a new possibility arising from the scalar
sector of Randall-Sundrum scenarios.
The gray region in both panels is again the theoretically
excluded region. The yellow region marked ``Tevatron Excluded'' makes
use of the latest Tevatron Data from the Higgs search \cite{D0CDF}. The most important channel for Higgs searches in the mass
range $150-200$ GeV is the process $p\bar{p}\to h_{SM}\to W^+W^-$ with the two
$W$'s decaying leptonically. One can then easily convert bounds on this
process into bounds in the Radion parameter space \cite{MTNO}, and as is seen in the
figure for $\Lambda_\phi=2$ TeV these bounds do reach our parameter space.
\section{Conclusions}

In this paper we have extended the study of Radion phenomenology with
matter fields in the Bulk to the case of Higgs and Radion mixing.
From the study of Higgs-Radion mixing in RS1 \cite{GRW,CGK,Rizzo,Gunion} one might think that
very similar results would hold modulo the relative
enhancements/suppressions observed when Fields are in the Bulk \cite{CHL}. 
This was actually the case for the $pp\to\phi\to ZZ$ channel but surprising
effects happened in the $\gamma\gamma$ channel. Of course in this
channel, both the production and the decay happen via loops and
perhaps it is not surprising that presumably benign changes in the underlying
model can produce significant phenomenological differences.
In particular we pointed out that a non negligible region of parameter
space allows for important di-photon signals even in a mass range well
above the $WW$ and $ZZ$ thresholds. In this case the main effect is
caused not because of a huge enhancement of the Radion coupling to
photons (although it is enhanced for large Radion masses) but because
of a suppression of the couplings to the SM massive particles. 
In the context of Higgs-Radion mixing, this signal could then play an 
important discriminating role between different models of warped extra
dimensions.

Other interesting features of the Radion with Bulk Fields pointed out
in \cite{CHL} such as its coupling to fermions depending directly on the 5D bulk 
mass parameters $c_{L,R}$ might also be enhanced by the effects of the
Higgs-Radion mixing and should also be looked at carefully although we
leave this for the future.

\section{Acknowledgments}

I would like to thank Kaustubh Agashe, Csaba Csaki, Daniele Dominici, Bohdan Grzadkowski,
Jack Gunion, Jay Hubisz and Nobuchika Okada for past and present discussions, as well as 
the KITP center for its hospitality. This research was supported in part by the National Science 
Foundation under Grant No. NSF PHY05-51164.


\end{document}